\documentclass[sigconf]{acmart}
\AtBeginDocument{%
  }
\setcopyright{acmcopyright}
\copyrightyear{2022}
\acmYear{2022}
\acmDOI{XXXXXXX.XXXXXXX}

\acmConference[ASE '22]{37th IEEE/ACM International Conference on Automated Software Engineering}{October 10--14, 2022}{Ann Arbor, MI, United States}
\acmPrice{15.00}
\acmISBN{978-1-4503-XXXX-X/18/06}

\begin{document}
\begin{sloppy}

\title{Taming Multi-Output Recommenders for Software Engineering}

\author{Christoph Treude}
\email{christoph.treude@unimelb.edu.au}
\affiliation{%
  \institution{The University of Melbourne}
  \city{Melbourne}
  \state{Victoria}
  \country{Australia}
}

\renewcommand{\shortauthors}{Treude}

\begin{abstract}
Recommender systems are a valuable tool for software engineers. For example, they can provide developers with a ranked list of files likely to contain a bug, or multiple auto-complete suggestions for a given method stub. However, the way these recommender systems interact with developers is often rudimentary---a long list of recommendations only ranked by the model's confidence. In this vision paper, we lay out our research agenda for re-imagining how recommender systems for software engineering communicate their insights to developers. When issuing recommendations, our aim is to recommend diverse rather than redundant solutions and present them in ways that highlight their differences. We also want to allow for seamless and interactive navigation of suggestions while striving for holistic end-to-end evaluations. By doing so, we believe that recommender systems can play an even more important role in helping developers write better software.
\end{abstract}



\maketitle

\section{Introduction}

A recommender system for software engineering is defined as ``a software application that provides information items estimated to be valuable for a software engineering task in a given context''~\cite{robillard2009recommendation}. These systems aim to assist developers in various activities from reusing code to writing effective bug reports. Recent years have witnessed the use of machine learning and deep learning in many recommender systems, making them more accurate and efficient~\cite{portugal2018use}.

In terms of output, recommender systems for software engineering are meant provide value by ``exposing users to the most interesting items, and by offering novelty, surprise, and relevance''~\cite{robillard2009recommendation}. What this description implicitly conveys is true in practice for many recommender systems for software engineering: their benefits are realised through their ability to recommend multiple items for a given task in a given context. For example:

\begin{itemize}
    \item Bug localisers provide a list of potentially buggy files ranked according to the system's confidence in the recommendation (e.g.,~\cite{thung2014buglocalizer}).
    \item Source code recommenders or synthesizers suggest multiple completions for the same source code context and task (e.g.,~\cite{campbell2017nlp2code}).
    \item Query reformulators provide a list of suggestions for better queries to allow more effective information retrieval (e.g.,~\cite{cao2021automated}).
    \item API recommenders provide a list of relevant API elements, such as methods, for a given scenario (e.g.,~\cite{chen2021holistic}).
    \item Change location recommenders suggest multiple relevant spots for supplementary bug fixes (e.g.,~\cite{xia2017effective}).
    \item Tag recommenders for Stack Overflow and similar sites recommend multiple tags per post (e.g.,~\cite{liu2018fasttagrec}).
    \item Recommenders for `who should fix this bug' provide multiple answers to this question~\cite{anvik2006should}.
\end{itemize}

The multi-output nature of recommender systems for software engineering is also reflected in the evaluation criteria that have been used in the research literature, such as the ratio of inputs for which at least one relevant result is returned within the top-$k$ results (Hits@K), the ranks of relevant results within a ranked list (mean average precision), or the multiplicative inverse of the rank of the first relevant result (mean reciprocal rank)~\cite{rahman2018improving}.

In the field of software engineering, there has been a plethora of work on improving the performance of recommender systems. However, surprisingly little research has been conducted on how these systems should communicate their insights to developers, especially in cases where the recommender system provides multiple recommendations for a given task in a given context. Many approaches in the literature are not accompanied by corresponding user interfaces, and in the few cases where a user interface is included, this is often limited to a list of recommendations ordered by the recommender's confidence in them. We argue that this does not do justice to the complexity of many software engineering tasks where multiple parallel recommendations need to be carefully compared and considered from multiple perspectives in order to understand complex solution spaces. Instead, many user interfaces ascribe too much importance to their top recommendation, often making navigation to further recommendations unnecessarily cumbersome. We show examples of this in our next section.

Re-imagining how multi-output recommenders for software engineering communicate their insights to developers can unlock the potential of recommender systems that have already been developed, as well as pave the way for new systems that explicitly cater to the multi-solution nature of software engineering tasks. The goals of our research agenda are: 

\begin{description}
\item[Diversification] Novelty and diversity have long been used as metrics for the evaluation of information retrieval systems in other fields~\cite{clarke2008novelty}. We argue that diversity also plays a crucial role when exploring the solution space for software engineering tasks---providing diverse recommendations promotes serendipity and creativity~\cite{ziarani2021serendipity} and can help resolve ambiguity and avoid redundancy.
\item[Representation] In many software engineering tasks, such as the synthesis of source code, small differences between recommendations completely change the semantics of the solution, as evidenced, for example, by small syntactic changes (such as reversing a logical operator) leading to large semantic changes~\cite{hayes2002applying}, or the large number of single-statement bugs in source code~\cite{karampatsis2020often}. We argue that recommender systems should visually support developers in comparing and contrasting multiple recommendations.
\item[Navigation] Recommender systems are often implemented as a separate window or tab from the main application, with multiple recommendations displayed at once. We argue for an integrated approach that enables requests for further recommendations that combine specific aspects of previous ones. This would allow developers to easily navigate between recommendations, similar to NLP2Code's `cycle-through' functionality~\cite{campbell2017nlp2code}. This kind of navigation support is essential to allow developers to adequately explore the solution space before deciding which recommendation to follow.
\item[Evaluation] Like other machine-learning-enabled systems, we argue that the evaluation of recommender systems for software engineering needs to include ``evaluating components individually (including the model) as well as their integration and the entire system, often including evaluating and monitoring the system online (in production)''~\cite{nahar2022collaboration}. In particular, we call for more research on how developers do or do not make use of multiple recommendations issued for a given task in a given context.
\end{description}

The remainder of this vision paper is structured as follows. We provide three motivating examples in Section~2 before we describe our research agenda in Section~3. Section~4 concludes this work.

\section{Motivating Examples}

In this section, we describe three motivating examples from GitHub Copilot\footnote{\url{https://copilot.github.com/}}, BugLocalizer~\cite{thung2014buglocalizer}, and SEQUER~\cite{cao2021automated}.

\subsection{GitHub Copilot}

\begin{figure}
\centering
\includegraphics[width=.8\linewidth]{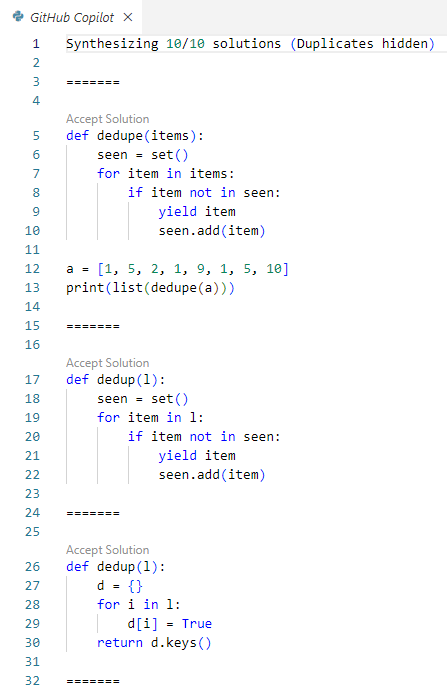}
\caption{Three suggestions by GitHub Copilot for \\ \texttt{\# deduplicate list} \\ \texttt{def} }
\label{fig:copilot}
\end{figure}

GitHub Copilot is an example of a source code recommender system built by training a deep learning model on millions of open source repositories: The source code of these repositories acts as training data, allowing the model to learn ``how to program''~\cite{ciniselli2022extent}. GitHub Copilot was released in October 2021 as a technical preview and can be installed as a Visual Studio Code extension.\footnote{\url{https://marketplace.visualstudio.com/items?itemName=GitHub.copilot}} Once installed, GitHub Copilot automatically suggests the code that it thinks the developer might want as the developer is typing.

While GitHub Copilot will always show its ``best recommendation'' in the editor,\footnote{\url{https://github.blog/2022-03-29-github-copilot-now-available-for-visual-studio-2022/}} developers can press \texttt{Ctrl + Enter} to view up to ten suggestions in a separate pane.\footnote{\url{https://github.com/github/copilot-docs/blob/main/docs/visualstudiocode/gettingstarted.md}} Figure~\ref{fig:copilot} shows a screenshot of this pane, containing three of the ten suggestions that GitHub Copilot issued for the following Python code:

\hspace{1cm}\texttt{\# deduplicate list}

\hspace{1cm}\texttt{def}

\noindent The first of these suggestions was automatically inserted into the source code as an inline suggestion and could be accepted by pressing the \texttt{Tab} key.

Considering the quality of the recommendations---all recommendations appear to produce the desired behaviour---the way in which the additional suggestions are communicated to a developer is surprisingly rudimentary. In terms of \textit{diversity}, the methods suggested in the first two recommendations are identical apart from the identifier names, but as the third recommendation shows, this is not the only way to complete the task. In terms of \textit{representation}, the differences between the first two recommendations are not immediately obvious: upon closer inspection, the first recommendation includes a method invocation with an example list, which is missing in the second recommendation. In terms of \textit{navigation}, interactions between different recommendations are not taken into account, e.g., it is not easily possible to choose the third recommendation but with the example and \texttt{print} statement from the first recommendation. 

\subsection{BugLocalizer}

\begin{figure}
\centering
\includegraphics[width=.8\linewidth]{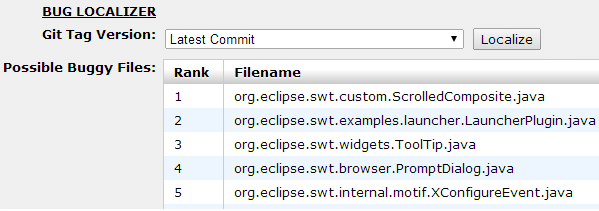}
\caption{BugLocalizer Main User Interface, from \cite{thung2014buglocalizer}}
\label{fig:buglocalizer}
\end{figure}

Bug localisation is another popular research area in the context of recommender systems for software engineering, defined as the process of identifying where to make changes in response to a bug report~\cite{moreno2014use}. Research efforts in this area are often evaluated in terms of performance metrics that take the multi-output nature of the problem into account (e.g., Hits@K~\cite{rahman2018improving}), while research on how to communicate the results of bug localisation to developers is rare. A notable exception is BugLocalizer~\cite{thung2014buglocalizer}, an extension of the Bugzilla issue tracking system aimed at disseminating research in bug localisation to practitioners. Figure~\ref{fig:buglocalizer} shows its interface after processing a bug report, indicating the ``top-$k$ files that are deemed the most likely to be buggy among files in the latest commit of the project’s git repository''~\cite{thung2014buglocalizer}. 

Similarly to the previous example, we argue that the way in which bug localisation results are currently communicated to developers does not unlock the full potential of the underlying models. Although the paths and names of the files in Figure~\ref{fig:buglocalizer} suggest \textit{diversity} at least within the \texttt{org.eclipse.swt} package, the \textit{representation} does not help to understand how these recommendations differ from each other, e.g., to what extent the changes in these files were similar and/or related. In terms of \textit{navigation}, it is not possible to navigate the recommendations along axes such as the call graph or package hierarchy, making the evaluation of each recommendation a tedious and manual task.

\subsection{SEQUER}

\begin{figure}
\centering
\includegraphics[width=.8\linewidth]{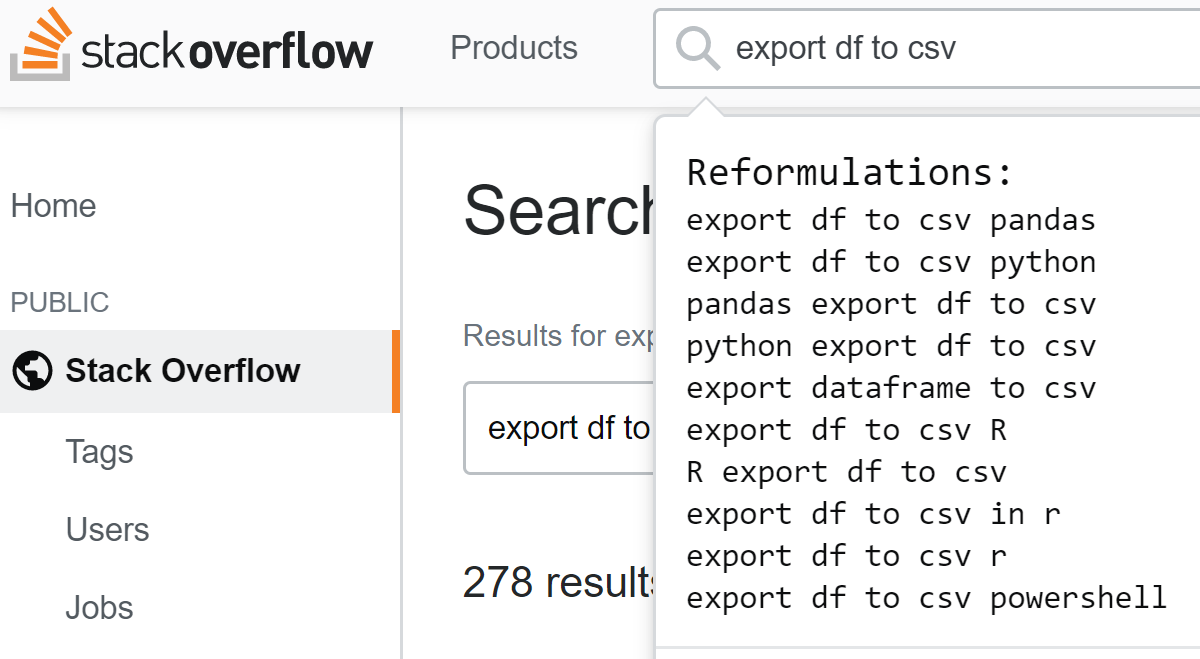}
\caption{SEQUER Browser Extension, from \cite{cao2021automated}}
\label{fig:sequer}
\end{figure}

Query reformulation is a popular research area in software engineering~\cite{haiduc2013automatic}, covering a wide range of developer tasks, from Stack Overflow search~\cite{cao2021automated} and search for software projects~\cite{li2016query} to concept location~\cite{rahman2016quickar} and bug localisation~\cite{florez2021combining}. In most of these scenarios, the research tools show a list of recommendations for reformulations to developers. Figure~\ref{fig:sequer} shows an example of SEQUER~\cite{cao2021automated}, a browser plugin that automatically analyses the query content on Stack Overflow and recommends the top 10 query reformulation candidates to developers.\footnote{\url{https://github.com/kbcao/sequer}} SEQUER was trained on the Stack Overflow activity logs and has been shown to be able to automatically correct misspelled terms, add language restrictions, remove overly specific query terms, replace symbols with their corresponding text, enclose domain-specific terms in double quotes, and simplify and refine Stack Overflow queries.

SEQUER outperformed its baselines in terms of metrics, such as the multiplicative inverse of the rank of the first relevant result (mean reciprocal rank), but we argue that the interface between the underlying model and developers has room for improvement. As Figure~\ref{fig:sequer} shows, reformulation recommendations can be \textit{diverse} in terms of characteristics such as programming languages, but this diversity is not well \textit{represented}. In the example, the first five recommendations are specific to Python, the next four are specific to the R programming language, and the last one concerns PowerShell. In addition, several of the recommendations are redundant since they only change the order of tokens, e.g., the first and third recommendations. We believe that SEQUER could make better use of the limited number of spots in its list of recommendations by favouring diverse recommendations over redundant ones. In terms of \textit{navigation}, the interface does not offer navigation along axes such as programming language, library, or API element. Like many similar tools, the \textit{evaluation} of SEQUER focused on the performance of the underlying model, without an end-to-end evaluation to investigate how developers make use of the tool's multiple outputs.

\section{Research Agenda}

In this section, we outline our envisioned methodology for addressing our research goals. These goals follow directly from our motivating examples, but we acknowledge that there are likely other goals worthy of investigation in this context.

\subsection{Diversification}

To address our goal of diversifying the output of multi-output recommenders for software engineering, we argue for the adoption and customisation of work in the area of diversity optimisation for other application domains, such as the evolution of diverse sets of images~\cite{neumann2018discrepancy}. Diversity optimisation is ``a new family of optimisation algorithms that, instead of searching for a single optimal solution to solving a task, searches for a large collection of solutions that all solve the task in a different way''~\cite{cully2019autonomous}. Such optimisations require an underlying numeric representation of the problem domain, similar to other software engineering tasks, such as vulnerability prediction, where artefacts must be represented as numeric vectors to be used as input for deep learning models~\cite{alon2019code2vec}. Such representations are commonly based on syntax or semantics, e.g., learning representations of source code at the level of tokens (e.g., abstract syntax trees) or statements (e.g, control flow graphs), with recent approaches combining low-level syntactic information from the local context and high-level semantic information from the global context into a single representation~\cite{jiang2022hierarchical}. Representation of text-based artefacts, such as file names or auto-complete suggestions, can be achieved through domain-specific word embeddings~\cite{efstathiou2018word}. We believe that the use of such techniques can increase the diversity of recommendations and enable developers to explore the entire solution space for their tasks. Similar approaches have been successful in the diversification of reply suggestions for instant messaging systems~\cite{deb2019diversifying} and emails~\cite{buschek2021impact} where more suggestions have been shown to be particularly useful for non-native speakers and in fostering creativity~\cite{singh2022hide}.

\subsection{Representation}

To improve the representation of multiple recommendations, we argue for integrating concepts from change visualisation in the context of software engineering into user interfaces of recommender systems. For example, representing source code commits by mixing text-based diff information with visual representation and metrics characterising the changes has been well received by developers~\cite{gomez2015visually}. A key difference between diff representation and representing multiple recommendation is cardinality---a diff concerns two versions (or three in the case of a three-way merge~\cite{ghiotto2018nature}) whereas recommender systems for software engineering tend to return dozens of results for a given task in a given context. To overcome this limitation, the adoption of variant graphs~\cite{schmidt2009data} from the text processing community can allow the representation of commonality between recommendations and differences between them. We envision the use of boldface to indicate commonality between recommendations, as well as pop-over comments for explanations of differences, similar to Casdoc~\cite{nassif2022casdoc}.

\subsection{Navigation}

To enable effective navigation of multiple recommendations, we argue for borrowing mechanisms from poker games. In any given poker game, players must make the decision of which cards to keep and which to discard~\cite{barboianu2007draw}. This analogy can be used to help developers navigate recommendations. For example, when given recommendations about files that might be buggy, we envision that developers are able to highlight parts of the file paths that they would like to `keep' before asking the recommender system to generate further recommendations. Similarly, in the scenario of source code synthesis, we envision developers able to highlight parts of the code that they would like to see again while discarding parts that are irrelevant. In the scenario shown in Figure~\ref{fig:copilot}, a developer might highlight the example list and \texttt{print} statement from the first recommendation but ask the system for a different algorithm to process the list. We also believe that an integrated `cycle-through' functionality that shows recommendations in their context rather than in a separate window or pane, similar to NLP2Code~\cite{campbell2017nlp2code} and NL2Code~\cite{xu2022ide}, will improve the navigation of multiple recommendations.

\subsection{Evaluation}

To confirm that addressing the research goals presented above does indeed improve how multi-output recommender systems for software engineering communicate their insights to developers, we argue for system and usability testing, similar to related work on the evaluation of machine-learning-enabled systems~\cite{nahar2022collaboration}. Although the software engineering research community has so far focused on evaluating model performance~\cite{lu2022towards}, we lack behind other fields in terms of real-world integration of these models~\cite{sendak2020real}. Performance metrics such as the ranks of relevant results within a list (mean average precision) are insufficient to determine whether a single recommendation or the interaction of multiple recommendations led developers onto the right path for solving their tasks. There is `no silver bullet' for many software engineering tasks~\cite{brooks1987no} and it is na\"{i}ve to assume that a single recommendation will be sufficient to solve a non-trivial software engineering task. We need to embrace the multi-output nature of recommender systems for software engineering and enable these systems to effectively communicate their insights. Participant observations or interviews would be suitable methods to assess whether we were able to achieve these goals, for example~\cite{sendak2020real}.

\section{Conclusion}

The models that power automated recommender systems for software engineering have seen impressive performance increases over the last few years. However, these advances are not always accompanied by adequate ways in which the systems communicate their insights to developers. This can lead to misunderstandings and frustration on the part of developers who may not be able to easily interpret or use the recommendations generated by these systems. In this vision paper, we lay out our research agenda for re-imagining how systems such as bug localisers, source code synthesisers, and API recommenders can enable developers to navigate the diverse solution spaces inherent in many software engineering tasks. Our research agenda calls for recommending diverse rather than redundant solutions, aligned with the `no silver bullet' nature of many software engineering tasks. We envision a representation of recommendations that enables developers to effortlessly spot similarities and differences, as well as interactions between multiple recommendations, and navigation mechanisms that allow developers to ask for further recommendations that contain aspects of items already recommended. To evaluate whether we are making progress towards these goals, we will require holistic end-to-end system and usability evaluations of recommender systems. We believe that this work will not only improve the effectiveness of recommender systems for software engineering but also help to build a community of developers who are confident in their ability to use automated tools for software engineering tasks.

\begin{acks}

The author thanks Larissa Salerno de Castro and the anonymous reviewers for their comments and suggestions.

\end{acks}

\bibliographystyle{ACM-Reference-Format}

\end{sloppy}
\end{document}